   \documentclass[prl,aps,twocolumn,showpacs]{revtex4}
\usepackage[dvips]{graphicx}
\usepackage{dcolumn}%
\usepackage{amsmath}%
\usepackage{amsfonts}%
\usepackage{amssymb}
\providecommand{\U}[1]{\protect\rule{.1in}{.1in}}
\newcommand{\be}{\begin{equation}}
\newcommand{\ee}{\end{equation}}
\newcommand{\bea}{\begin{eqnarray}}
\newcommand{\eea}{\end{eqnarray}}
\newcommand{\bt} {\begin{tabular}}
\newcommand{\et} {\end{tabular}}
\newcommand{\nn}{ \nonumber}
\newcommand{\ds}{\displaystyle}
\newcommand{\ba} {\begin{array}}
\newcommand{\ea} {\end{array}}
\begin{document}

\title{The effect  of dephasing on thermoelectric efficiency of molecular junctions}

\author{Natalya A. Zimbovskaya}

\affiliation
{Department of Physics and Electronics, University of Puerto 
Rico-Humacao, CUH Station, Humacao, Puerto Rico 00791, USA; \\
 Institute for Functional Nanomaterials, University of Puerto Rico, San Juan, Puerto Ruco 00931, USA} 

\begin{abstract}
In this work we report the results of theoretical analysis of the effect of thermal environment on the thermoelectric efficiency of molecular junctions. The environment is represented by two thermal phonon baths associated with the electrodes which are kept at different temperatures. The analysis is carried out using the Buttiker model within the scattering matrix formalism to compute electron transmission through the system. This approach is further developed, so that the dephasing parameters are expressed in terms of relevant energies including the thermal energy, strengths of coupling between the molecular bridge and the electrodes and characteristic energies of electron-phonon interactions. It is shown that the latter significantly affect thermoelectric efficiency by destroying coherency of the electron transport through the considered system.
   \end{abstract}

%\pacs{72.15.Gd,71.18.+y}%%{71.18.+y, 71.20-b, 72.55+s}

\date{\today}
\maketitle

\section{i. introduction}

For the past three decades, significant efforts were applied to study thermal and thermoelectric transport in mesoscopic and nanoscale systems  of various kinds including quantum dots and/or molecules  attached to conducting electrodes. The latter serve as source and drain reservoirs for traveling charge carriers. Here, we concentrate on the analysis of thermoelectric properties of these systems. Below they are referred to as thermoelectric junctions.
In part, the research interest appeared because thermoelectric junctions are expected to be useful in building up highly efficient energy conversion devices.  Also, studies of thermoelectric properties of these systems can result in a deeper insight into the nature of general transport mechanisms and bring additional information on the electronic and vibrational excitation spectra of molecules \cite{1,2,3,4,5,6,7,8,9,10,11,12,13,14,15}.  Presently, it is established that thermoelectric properties of quantum dots and/or molecules may be strongly affected by Coulomb interactions between charge carriers \cite{16,17,18,19,20,21,22}. Coulomb interactions lead to violation of the Wiedemann-Franz law in nanoscale thermoelectric junctions thus providing an enhancement of thermoelectric efficiency of these systems \cite{21,23}. The thermal efficiency may be further increased due to the influence of quantum interference effects which may strongly affect electron transport characteristics  \cite{9,24,25,26,27,28,29}.

In  general studies of thermoelectric transport through molecules, quantum dots and similar systems  one must imply that both atomic vibrations and charge carriers contribute to the energy transfer. Therefore, a unified description of the electrons and phonons dynamics is needed to thoroughly analyze thermoelectric properties of molecular junctions. The means for such analysis are provided by the nonequilibrium Green's functions formalism (NEGF), as described in the review \cite{30}, and some other works (see e.g. \cite{13,31,32,33,34,35}). However, application of this formalism to realistic models simulating thermoelectric junctions is extremely difficult. Several simplified approaches were developed and used to analyze the specifics of heat transfer and other related phenomena in quantum dots and molecules taking into account  the contribution of phonons and electron-phonon interactions \cite{2,13,36,37,38}. Nevertheless, these studies are not completed so far. 

The phonons contributing to the charge and energy transfer may be subdivided in two classes: vibrational phonons associated with molecular vibrations and thermal phonons associated with random nuclear motions in the environment. In the present work we aim at theoretical analysis of the effect  of thermal phonons on the thermoelectric characteristics of molecules and other similar systems. To carry on this analysis we combine NEGF with  the approach first suggested by Buttiker to describe quantum transport through molecules \cite{39}. An important advantage of this approach is that it could be easily adapted to analyze  various aspects of incoherent/inelastic transport through molecules (and some other mesoscopic systems) avoiding complicated and time-consuming methods based on more advanced formalisms.

\section{ii. Main equations}

For simplicity, in the following computations we simulate a molecule/quantum dot by a single level with the energy $ E_0. $ We assume that this single-level bridge is coupled to the pair of dissipative reservoirs, as shown in the Fig. 1.  While on the bridge, an electron could be scattered into one of the reservoirs through the channels 3 and 4 (or 5 and 6) with a certain probability. In the reservoir, it undergoes inelastic scattering accompanied by phase breaking, and afterwards returns to the bridge with the same probability. In the present analysis, the reservoirs are treated as phonon baths representing thermal phonons associated with the left and right electrodes. Within the accepted model we imply that there is no phonon thermal conductance through the junction. This seems a reasonable assumption for experiments give low values of phonon thermal conductance in several thermoelectric molecular junctions \cite{40,41}. This may be attributed to the fact that in many molecules the majority of vibrational transitions lie above the range determined by thermal energy when temperature takes on values of the order of or lower than the room temperature \cite{14,29}.
Within the Buttiker model, the electron transport through a thermoelectric junction is considered as combination of tunnelings through potential barriers separating the electrodes from  the molecule/quantum dot and interaction with the reservoirs coupled to the bridge site.

\begin{figure}[t] %%% fig. 1
\begin{center}
\includegraphics[width=5.7cm,height=8.8cm,angle=-90]{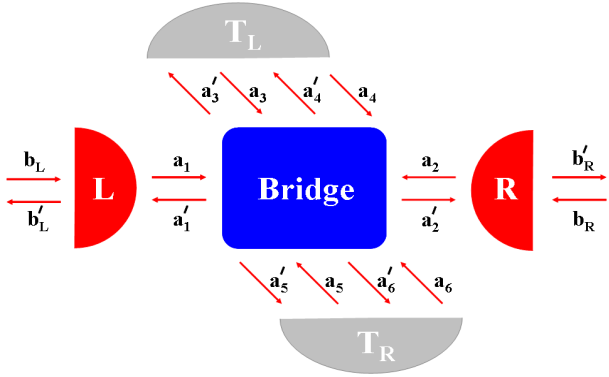}
\caption{(Color online) Schematics of the considered system. Semicircles  represent the left and right electrodes, square stands for the molecule/quantum dot sandwiched in between.  Dephasing/dissipative reservoirs are associated with the electrodes and characterized by temperatures $ T_L $ and $ T_R, $ respectively.
}
 \label{rateI}
\end{center}\end{figure}

The Buttiker approach was applied to describe and analyze electron transport through molecules in several works (see e.g. Refs. \cite{42,43,44}. Following this approach, one can present particle fluxes $ J_i' $  outgoing from the system as linear combinations of incoming fluxes $ J_k $ where the indexes $ i,k $ label the channels for transport. For the adopted model $ 1 \leq i \leq k \leq 6. $
\be 
J_i' = \sum_k T_{ik} J_k  \label{1}
\ee
In these equations, the coefficients $ T_{ik} $ are related to  matrix elements of the scattering matrix $ M $ namely: $ T_{ik} = |M_{ik}|^2. $ The scattering matrix expresses outgoing wave amplitudes $ b_L',b_R',a_3',a_4',a_5',a_6' $ in terms of incident ones $ b_L,b_R,a_3,a_4,a_5,a_6. $ To provide the charge conservation in the system, zero net current should flow in the channels linking the  bridge site with the dephasing reservoirs, so we may write the following equations:
\begin{align}
J_3 + J_4 - J_3' - J_4' = 0 ,\nn
\\ 
J_5 + J_6 - J_5' - J_6' = 0 . \label{2}
\end{align} 
To find the expressions for the matrix elements $ M_{ik} $ we first consider a subsystem including the left electrode with the associated dephasing reservoir and the bridge site. The expression for the matrix $s^{(1)} $ relating the wave amplitudes $a_1', a_2', a_3', a_4' $ to the wave amplitudes $a_1, a_2, a_3, a_4 $ has the form \cite{39}:
\be 
s^{(1)} = \left( \ba{cccc}
0  & \sqrt{1 - \epsilon_L} & \sqrt {\epsilon_L} &  0
\\
 \sqrt{1 - \epsilon_L} & 0  & 0 & \sqrt {\epsilon_L}
\\
 \sqrt{ \epsilon_L} & 0  & 0 & - \sqrt {1-\epsilon_L}
\\
0  & \sqrt{\epsilon_L} & -\sqrt {1-\epsilon_L} &  0 
\\
\ea  \right) \label{3}
\ee
where the phenomenological scattering probability $ \epsilon_L $ corresponds to the reservoir associated with the left electrode. Also, an electron tunneling through a single potential barrier separating  this electrode from the molecule/quantum dot is characterized by the transmission and reflection amplitudes $(t_L$ and $r_L,$ respectively). These are the matrix elements of a $ 2\times 2 $ matrix:
\be
s_L = \left( \ba{cc}
t_L & r_L \\  r_L & t_L \ea \right)  . \label{4}
\ee
Combining Eqs. (\ref{3}) and (\ref{4}) one obtains the expression for the scattering matrix $ M^{(1)} $ relating $b_L', a_2', a_3', a_4' $ to  $b_L, a_2, a_3, a_4 $:
\be
M^{(1)} = \left(\ba{cccc}
r_L & \alpha_Lt_L & \beta_Lt_L & 0 
\\
\alpha_Lt_L & \alpha_L^2r_L & \alpha_L\beta_Lr_L & \beta_L
\\\beta_Lt_L & \alpha_L\beta_Lr_L & \beta_L^2r_L & -\alpha_L
\\
0 & \beta_L & -\alpha_L & 0  \ea \right).  \label{5}
\ee
Here, $ \alpha_L = \sqrt{1 - \epsilon_L},\ \beta_L = \sqrt{\epsilon_L}. $

Now, we take into consideration the remaining elements of the original system. The matrix $ M^{(2)} $ which relates $a_2', b_R', a_5', a_6' $ to  $a_2, b_R, a_5, a_6 $ is \cite{43,44}:
\be
M^{(2)} = \left( \ba{cccc}
\alpha_R^2 r_R & \alpha_R t_R & \beta_R & \alpha_R \beta_R r_R 
\\
\alpha_R t_R & r_R & 0 & \beta_R t_R
\\
\beta_R & 0 & 0 & -\alpha_R
\\
\alpha_R \beta_R r_R & \beta_R & - \alpha_R & \beta_R^2 r_R
\ea \right) \label{6}
\ee
where $ \alpha_R = \sqrt{1 - \epsilon_R},\ \beta_R = \sqrt{\epsilon_R}, $ the scattering probability $ \epsilon_R $ is associated with the right reservoir, and the transmission $(t_R) $ and reflection $(r_R) $ amplitudes characterize electron tunneling through the potential  barrier between the molecule (bridge) and the right electrode. Using Eqs. (\ref{5}) and (\ref{6}) and excluding the wave amplitudes $a_2,a_2' $ which correspond to the transport inside the system, we get the following expression for the scattering matrix:

\begin{widetext}
\be 
M = \frac{1}{Z}  \left \{\ba{cccccc}
r_L + \alpha_L^2\alpha_R^2r_R  &\alpha_L \alpha_R t_L t_R   &
\beta_L t_L    &  \alpha_L\alpha_R^2 \beta_L t_L r_R   &
\alpha_L  \beta_R t_L   &  \alpha_L \alpha_R \beta_R t_L r_R  
\\ 
\alpha_L \alpha_R t_L t_R   &  r_R + \alpha_L^2 \alpha_R^2 r_L   &
\alpha_L \alpha_R \beta_L t_R r_L   &  \alpha_R \beta_L t_R &  
\alpha_L^2 \alpha_R \beta_R t_R r_L  &  \beta_R t_R
\\ 
\beta_L t_L  & \alpha_L \alpha_R \beta_L t_R r_L  &
\beta_L^2 r_R  & \alpha_L(\alpha_R^2 r_L r_R -1) &
\alpha_L \beta_L \beta_R r_L  &  \alpha_L \alpha_R \beta_L \beta_R r_L  r_R
 \\ 
\alpha_L\alpha_R^2 \beta_L t_L r_R  & \alpha_R \beta_L t_R &
\alpha_L(\alpha_R^2 r_L r_{R}-1) & \alpha_R^2 \beta_L^2 r_R &   \beta_L \beta_R   &  \alpha_R \beta_L \beta_R r_R 
 \\
\alpha_L \beta_R t_L  & \alpha_L^2 \alpha_R \beta_R t_R r_L &
\alpha_L \beta_L \beta_R r_L  & \beta_L \beta_R &
\alpha_L^2 \beta_R^2 r_L  & \alpha_R (\alpha_L^2 r_L r_R -1) 
\\
\alpha_L \alpha_R \beta_R r_R t_L & \beta_R t_R &
\alpha_L\alpha_R \beta_L \beta_R r_L r_R & \alpha_R \beta_L \beta_R r_R & \alpha_R(\alpha_L^2 r_L r_R-1) & \beta_R^2 r_R \\
\ea \right \} .  \label{7}
\ee
\end{widetext}
Here, $ Z = 1 - \alpha_L^2 \alpha_R^2 r_L r_R. $

Solving the system of linear equations (\ref{1}), (\ref{2}) one obtains the following expression for the electron transmission $ T(E) $ which coincides with the corresponding result reported by D'Amato and Pastawski \cite{42}:
\be
T(E) = \frac{J_2'}{J_1} = T_{21} + \sum_{i,j} K_i^{(2)} (W^{-1})_{ij} K_j^{(1)} .
 \label{8}  \ee
Within the considered model, $ 1 \leq i,j \leq 2, $
\begin{align}
K_i^{(1)} = T_{2i+1,1} + T_{2i+2,1},
\nn \\ 
K_i^{(2)} = T_{2,2i+1} + T_{2,2i+2}, \label{9}
\end{align}
and $W^{-1} $ is the matrix inversed with respect to $ 2\times 2 $ matrix $ W , $ whose matrix elements are given by:
\be
W_{ij} = (2 - R_{ii}) \delta_{ij} - \tilde R_{ij} (1 - \delta_{ij}). \label{10}
\ee
In this expression, the following denotations are used:
\begin{align} &
R_{ii}= T_{2i+1, 2i+1} + T_{2i+2, 2i+2} + T_{2i+2, 2i+1} + T_{2i+1, 2i+2},
\nn\\ &
\tilde R_{ij} = T_{2i+1, 2j+1} + T_{2i+1, 2j+2} + T_{2i+2, 2j+1} + T_{2i+2, 2j+2}.           \label{11}
\end{align}

Assuming that both dephasing reservoirs are detached from the  bridge $(\epsilon_L = \epsilon_R = 0), $ the transport through the system becomes coherent and elastic. In this case, the electron transmission given by Eqs. (\ref{8})-(\ref{11}) is reduced to a simple form: 
\be
T(E) = \frac{t_L^2t_R^2}{(1 + r_Lr_R)^2}. \label{12}
\ee 
As known, the expression for the electron transmission in the case of coherent transport may be presented as follows:
\be
T(E) \equiv  g^2(E) = Trace \big[(\Gamma_{L\sigma}(E) G^r_\sigma(E) \Gamma_{R\sigma}(E) G^a_\sigma (E)\big]                         \label{13}
\ee
where $ G^{r,a}_\sigma (E)$  are the retarded and advanced Green's functions associated with the molecule/quantum dot bridging the electrodes, and self-energy terms $\Gamma_{L,R;\sigma} $ describe the coupling between the electron of a certain spin orientation on the bridge and the corresponding electrode. For a symmetrically coupled system $(\Gamma_{L\sigma} = \Gamma_{R\sigma} = \Gamma),$ the expression for  electron transmission may be reduced to the form: 
\be
 T(E) = \frac{i}{2}\Gamma(E) \sum_\sigma\big[G_\sigma^r(E) - G_\sigma^a(E) \big].  \label{14}
\ee
Provided that electron transport through the system is undisturbed  by electron-phonon interactions, and disregarding spin-flip processes, the retarded Green's function $ G_\sigma^r(E) $ may be approximated as  \cite{45}:
\be
G_\sigma^r(E) = \frac{E - E_0 - \Sigma_2^\sigma - U\big(1 - \big<n_{-\sigma}\big>\big)}{\big(E - E_0 - \Sigma_{0\sigma}\big)\big(E - E_0 - U - \Sigma_2^\sigma\big) + U\Sigma_{1\sigma}}. \label{15}
\ee 
Here,  $ U $ is the charging energy associated with Coulomb repulsion between the electrons on the molecular bridge/quantum dot and $ \big<n_{\pm\sigma}\big> $ are one-particle occupation numbers:
\be
 \big<n_{\sigma}\big> = \frac{1}{2\pi} \int dE \mbox{Im} \big[G_\sigma^<(E) \big] \label{16}
\ee
where $ G_\sigma^<(E) $ is the lesser Green's function for electrons on the bridge. 
  Self-energy corrections $ \Sigma_{0\sigma},\ \Sigma_{1\sigma},\ \Sigma_{2\sigma} $ appear in the expression for $ G_\sigma^r $ due to the coupling of the bridge to the electrodes. For example:
\be
\Sigma_{0\sigma}   = \sum_{r\beta} \frac{|\tau_{r\beta\sigma}|^2}{E - \epsilon_{r\beta\sigma} + i\eta} \equiv \Sigma_{0\sigma}^L +  \Sigma_{0\sigma}^R.                     \label{17}
\ee
In this expression, $ \epsilon_{r\beta\sigma} $ are  single-electron energies on the electrode $ \beta\ (\beta \in L,R), \ \tau_{r\beta\sigma} $ are coupling parameters characterizing the coupling of the electron states on the bridge to the electrodes and $ \eta $ is an infinitesimal positive parameter. These self-energy terms are closely related to the previously introduced  coupling strengths $ \Gamma_{\beta\sigma}, $ namely: $ \Gamma_{\beta\sigma} (E) = - 2\mbox{Im} \Sigma_{0\sigma}^\beta. $ The expressions (\ref{14})-(\ref{16}) were repeatedly employed in studies of thermal transport through quantum dots (see e.g Ref. \cite{26,46}).

For  a symmetrically coupled system, one may assume that the potential barriers separating the electrodes from the bridge are identical:$ t_L = t_R = t,\ r_L = r_R = r. $ Then the transmission amplitude could be easily expressed in terms of the corresponding Green's functions:
\be
t^2 = \frac{2g}{1 + g}. \label{18}
\ee

Within the Buttiker approach, the scattering probabilities $ \epsilon_{L,R} $ are introduced as phenomenological parameters. However, these parameters may be given an explicit physical meaning by expressing them in terms of the relevant energies. In the considered
 system, dissipation and loss of coherency appear due to the interaction of charge carriers with thermal phonons associated with the electrodes and represented by the dephasing reservoirs. Therefore, as was suggested in an earlier work \cite{47}, one can approximate these parameters as follows: 
\be
\epsilon_\beta = \frac{\Gamma_{ph}^\beta}{2(\Gamma_L + \Gamma_R) + \Gamma_{ph}^\beta} .  \label{19}
\ee
Here, $ \Gamma_{ph}^\beta$ represents the self-energy term originating from electron-phonon interactions occurring in the reservoir associated with the left/right electrode. Using NEGF and computing the relevant electron and phonon Green's functions within the self-consistent Born approximation, one can arrive at a relatively simple expression for $ \Gamma_{ph}^\beta $ \cite{30}:
\begin{align}
\Gamma_{ph}^\beta (E) = &\, 2\pi \lambda^2_\beta \int_0^\infty d\omega \rho_{ph}^\beta (\omega)
\nn\\  & \times \big\{
N(\omega) \big[\rho_{el}(E - \hbar\omega) + \rho_{el}(E + \hbar\omega) \big]  
\nn\\ & +
\big[1 - n(E - \hbar\omega)\big] \rho_{el}(E - \hbar\omega)
\nn \\ & +
n(E + \hbar\omega) \rho_{el}(E + \hbar\omega) \big\}. \label{20}
 \end{align}
In this expression, $ \rho_{el}(E) $ and $ n(E) $ are respectively the electron density of states associated with the bridge level and its steady state occupation, and $ \rho_{ph}^\beta(\omega) $ is the phonon spectral function  for the corresponding reservoir. We assume that the electrodes may be kept at different temperatures $T_\beta, $ so we introduce phonon distribution functions $ N_\beta (\omega) = \big\{\exp\big[\hbar\omega/k_BT_\beta\big] - 1\big\}^{-1} $ where $ k_B $ is the Boltzmann constant.
Finally, the constant $ \lambda_\beta $ characterizes the coupling strength for electron interactions with the thermal phonons belonging to the bath $\beta. $ 

The particular form of the phonon spectral functions $ \rho_{ph}^\beta (\omega) $ may be found basing on the molecular dynamic simulations. However, to qualitatively analyze the effect of dephasing on the thermoelectric transport, one may employ the approximation \cite{48}:
\be
\rho_{ph}^\beta(\omega) = \rho_{0\beta}\left(\frac{\omega}{\omega_{c\beta}}\right) \exp\left[-\frac{\omega}{\omega_{c\beta}}\right] \label{21}
\ee 
where the parameter $ \rho_{0\beta} $ is related to the electron-phonon coupling strength, and $ \omega_{c\beta} $ characterizes the relaxation time for the thermal phonons.

The electron density of states includes self-energy corrections which appear due to electron-phonon interactions. Therefore, Eq. (\ref{20}) is an integral equation for $\Gamma_{ph}^\beta. $ Substituting  the approximation (\ref{17}) into this equation, one may see that the major contribution to the integral over $ \omega $ originates from the region where $ \omega \ll \omega_c. $ Omitting the terms $ \hbar\omega $ in the arguments of all slowly varying terms in the integrand, we may reduce Eq. (\ref{20}) to the form:
\be
\Gamma_{ph}^\beta = \rho_{el} \big(E,\Gamma_L,\Gamma_R,\Gamma_{ph}^L,\Gamma_{ph}^R\big) \cdot Q(\lambda_\beta,\omega_{c\beta},T_\beta)  \label{22}
\ee
where  
\be
Q(\lambda_\beta,\omega_c,T_\beta ) = \frac{4\pi\lambda_\beta}{\hbar\omega_c}(k_B T_\beta)^2 \zeta \left(2; 1 +\frac{ k_B T_\beta}{\hbar\omega_{c\beta}} \right)  \label{23}
\ee
and $ \zeta(x;q) $ is the Riemann's $\zeta $ function.

The suggested approach gives means to theoretically analyze the effect of thermal phonons  on the thermoelectric properties of thermoelectric junctions. Using the obtained results given by Eqs. (\ref{7})-(\ref{21}), one may compute electron transmission implying that the difference in the  temperatures $ T_L $ and $ T_R $ can take on an arbitrary value. Therefore, these results may be employed to study thermoelectric properties of the  considered systems beyond the linear regime. As known, nonlinear thermoelectric properties of molecular junctions and similar systems presently attract significant interest \cite{33,37,46,49,50}.      
%%    Coulomb interactions between electrons on the molecule/quantum dot may accounted for by using appropriate expressions for the electron Green's functions incorporated into the expression for the coherent transmission (\ref{13}) as well as into the expression for the electron density of states$ \rho_{el}.$  
  However, in studies of thermoelectric characteristics of such systems beyond linear regime, one inevitably encounters a nontrivial task of introducing  and defining the local temperature for the bridge which differs from temperatures $ T_{L,R} $ associated with the electrodes. The definition of local temperature and related problems are thoroughly discussed in the recent review \cite{49}.

In the present work we avoid these difficulties by restricting further analysis with the linear temperature and bias regime. Also, we remark again that within the considered model the thermal conductivity associated with phonons is omitted for we do not include into consideration vibrational modes coupled to the bridge. Therefore, we may employ the following commonly used expressions for measurable thermoelectric characteristics:
\begin{align}  
S &  = - \frac{1}{eT} \frac{L_1}{L_0},  \label{24}
\\ 
ZT & = \frac{S^2GT}{\kappa} = \frac{L_1^2}{L_0L_2 - L_1^2}.  \label{25}
\end{align}
Here, $ G $ and $ \kappa $ are electron electrical and thermal conductances, $ S $ is the thermopower (Seebeck coefficient) and $ ZT $ is the dimensionless  thermoelectric figure of merit characterizing the efficiency of charge-driven cooling devices and/or heat-driven current generators. In deriving these expressions, it is assumed that $ T_R = T $ and $ T_L = T + \Delta T\ (\Delta T \ll T). $ The integrals $ L_n $ included in Eqs. (\ref{24}), (\ref{25}) are given by:
\be
L_n = \int(E - \mu)^n T(E) \frac{\partial f}{\partial E} dE \label{26}
\ee
where $ f $ is the Fermi distribution function for the energy $ E, $ and the chemical potential $ \mu $ characterizes the electrodes at zero bias.  Coulomb interactions between electrons on the molecule/quantum dot may be accounted for by using appropriate expressions for the electron Green's functions incorporated into the expression for the coherent transmission (\ref{13}) as well as into the expression for the electron density of states$ \rho_{el}.$  %%For a while, we omit from consideration the influence of Coulomb interactions assuming thet $ U = 0. $ 

\section{iii. Results and discussion}

Specific thermoelectric properties of the considered systems depend on the relation of four relevant energies. These are the strength of coupling of the bridge to the electrodes $ \Gamma, $ the electron-phonon coupling strength $ \lambda, $ the charging energy $ U $ characterizing  Coulomb interactions of electrons on the bridge, and the thermal energy $ k_B T. $ It was established that the greater values of $ ZT $ could be achieved in weakly coupled systems where the condition $ \Gamma \ll k_BT $ may be satisfied at reasonably low temperatures (see e.g. Ref. \cite{21}), so in further analysis we assume that considered system complies with this condition.

Also,  we assume that the considered quantum dot/molecule is symmetrically coupled to the electrodes $ (\Gamma_L = \Gamma_R = \Gamma) $ and two thermal baths are identical $(\omega_{cL} = \omega_{cR} = \omega_c,\ \lambda_L = \lambda_R = \lambda). $ Omitting for a while Coulomb interactions, one may derive a simple Lorentzian  expression for the electron density of states:
\be
\rho_{el} = \frac{1}{2\pi} \frac{\Gamma}{(E - E_0)^2 + (\Gamma + \Gamma_{ph})^2} . \label{27}
\ee
where $ \Gamma_{ph} = \Gamma_{ph}^L + \Gamma_{ph}^R. $
Substituting this expression into  Eq. (\ref{22}), we may solve this equation and arrive at a reasonable asymptotic expression for $ \Gamma_{ph}: $
\be
\Gamma_{ph} = \frac{\Gamma\delta^2(1 + \sqrt{1 + \delta^2})}{(E - E_0)^2 +  (1 + \sqrt{1 + \delta^2})^2 }   \label{28}
\ee
where $ \delta^2 = 2Q (\lambda,\omega_c,T)/\Gamma. $ Using this result and the expression (\ref{19}) for the scattering probabilities, we obtain:
\be
\epsilon_L = \epsilon_R =
\epsilon = \frac{1}{2} \frac{\delta^2(1 + \sqrt{1 + \delta^2})}{\ds \left(\frac{E - E_0}{\Gamma}\right)^2 + \frac{1}{2}\big(1 + \sqrt{1 + \delta^2}\big)^3}. \label{29}
\ee

The  parameter  $\epsilon $ values vary between 0 and 1. When $ \epsilon = 0, $ the bridge is detached from the reservoirs, and the electron transport is completely coherent and elastic. Within the opposite limit $ (\epsilon = 1) $ the transport is characterized by the overall phase randomization typical for inelastic sequential hopping. Within the adopted approach, the scattering probabilities depend on tunnel energy $ E. $ As well as electron transmission function $ T(E),$ they reach their maximum values $( \epsilon_{max} $ and $ T_{max} ,$ respectively) at $ E = E_0. $ This is shown in the left panels of the Fig. 2.

\begin{figure}[t] %%% fig. 2
\begin{center}
\includegraphics[width=8.8cm,height=4.5cm]{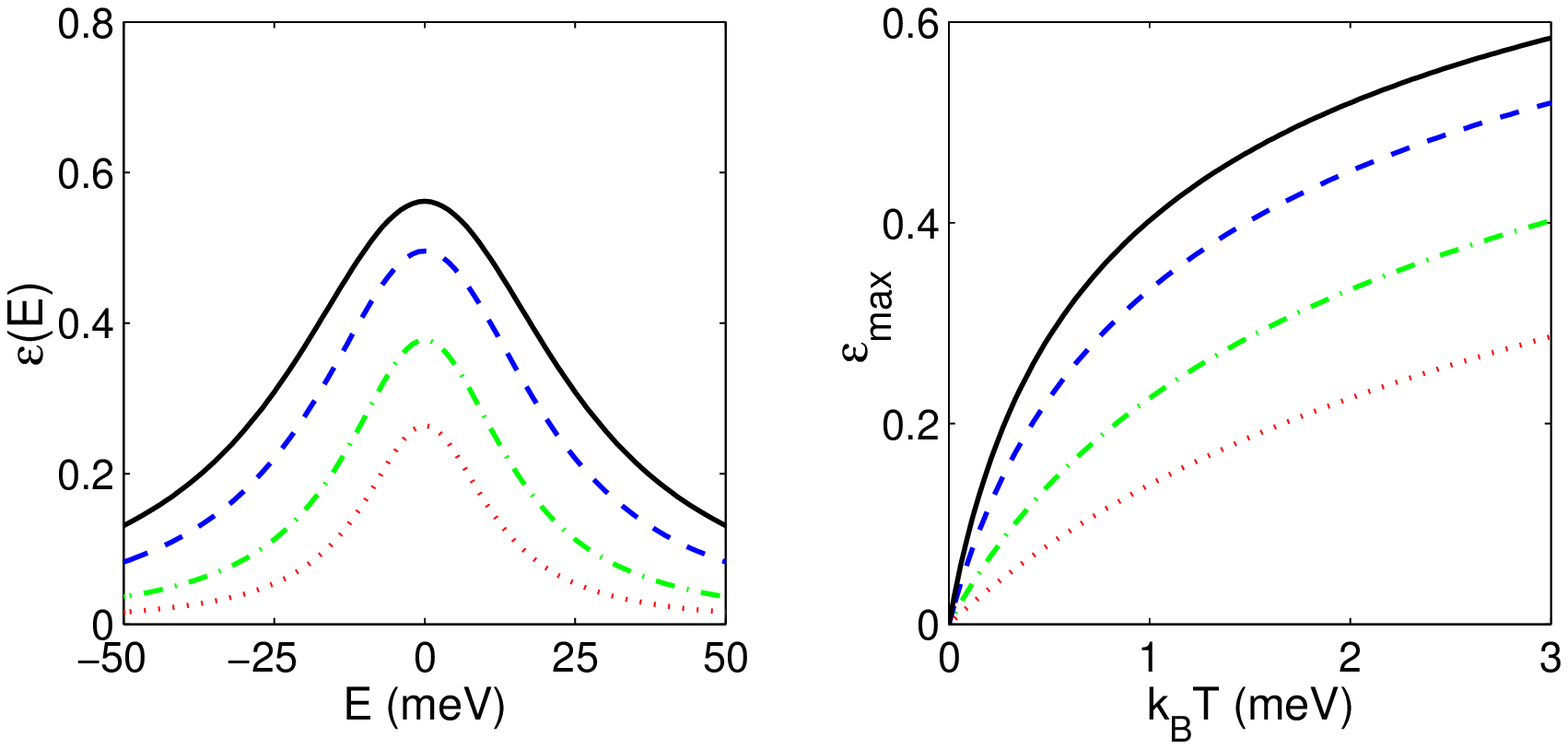}
\includegraphics[width=8.8cm,height=4.5cm]{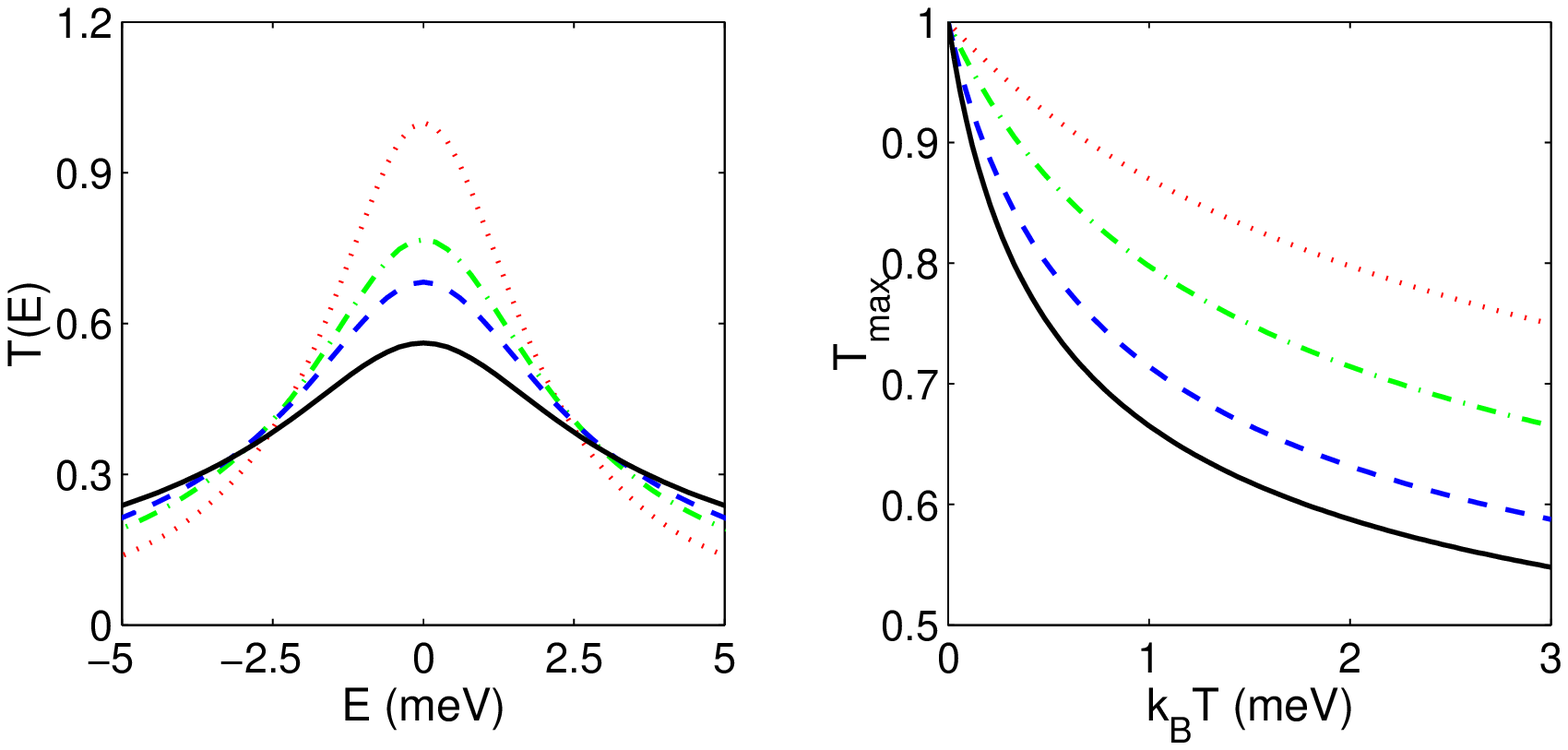}
\caption{(Color online) Left panels: Scattering probability (top) and electron transmission (bottom) as functions of tunnel energy $ E. $ Right panels: Temperature dependencies of peak values of $ \epsilon $ (top) and the electron transmission (bottom). All curves are plotted for a symmetrically coupled system $(\Gamma_L = \Gamma_R = \Gamma) $ with identical dephasing reservoirs assuming that $ T_L = T_R = T,\ \Gamma = 1 meV,\ E_0 = 0, $  $ \lambda = 1.5 meV $ (dotted line), $ \lambda = 3meV $ (dash-dotted line), $ \lambda = 6meV $ (dashed line), $ \lambda = 9meV $ (solid line).
In the left panels $ k_BT = 2.6meV.  $ 
}
 \label{rateI}
\end{center}\end{figure}

As follows from Eq. (\ref{29}), the character of the electron transport is determined by the value of the dimensionless parameter $ \delta. $ Transport remains nearly coherent when $ \delta \ll 1. $ On the contrary, the strong dephasing/dissipation occurs when $ \delta $ takes on values significantly greater than 1. To find a suitable estimate for $ \delta, $ one needs to approximate the Riemann's $\zeta $ function included into expression (\ref{23}). The approximation depends of the relation between the energies $ \hbar\omega_c $ and $ k_B T. $ As discussed in an earlier work \cite{43}, the effect of the thermal bath on the electron transport is significantly more pronounced when the lifetime of thermal excitations is sufficiently long $(\hbar\omega_c \ll k_B T),$ Under this condition, one may apply the estimation $ Q \approx 4k_B T\lambda. $ Correspondingly, $ \delta^2 \approx 4k_BT\lambda/\Gamma^2. $ This shows that the maximum value of the scattering probabilities $ \epsilon_{max} $  is determined with two parameters, namely, $T $ and $ \lambda. $ 
We remark that in the absence of electron-phonon interactions $(\lambda = 0),\ \epsilon \equiv \epsilon_m = 0 $ regardless of the energy $ E $ value, and $ T_{max} = 1.$ In general, 
 $ \epsilon_{max} $ increases when the temperature rises, and it takes on greater values when the electron-phonon interactions are getting stronger as illustrated in the Fig. 2. The enhancement of $\epsilon_{max} $ is accompanied by the decrease of maximum value of electron transmission $T_{max}. $ These results have an obvious physical sense because in the considered situation the phase randomization is inseparable from inelastic scattering of electrons by thermal phonons hindering  electron transport through the system.

\begin{figure}[t] %%% fig. 3
\begin{center}
\includegraphics[width=8.8cm,height=4.5cm]{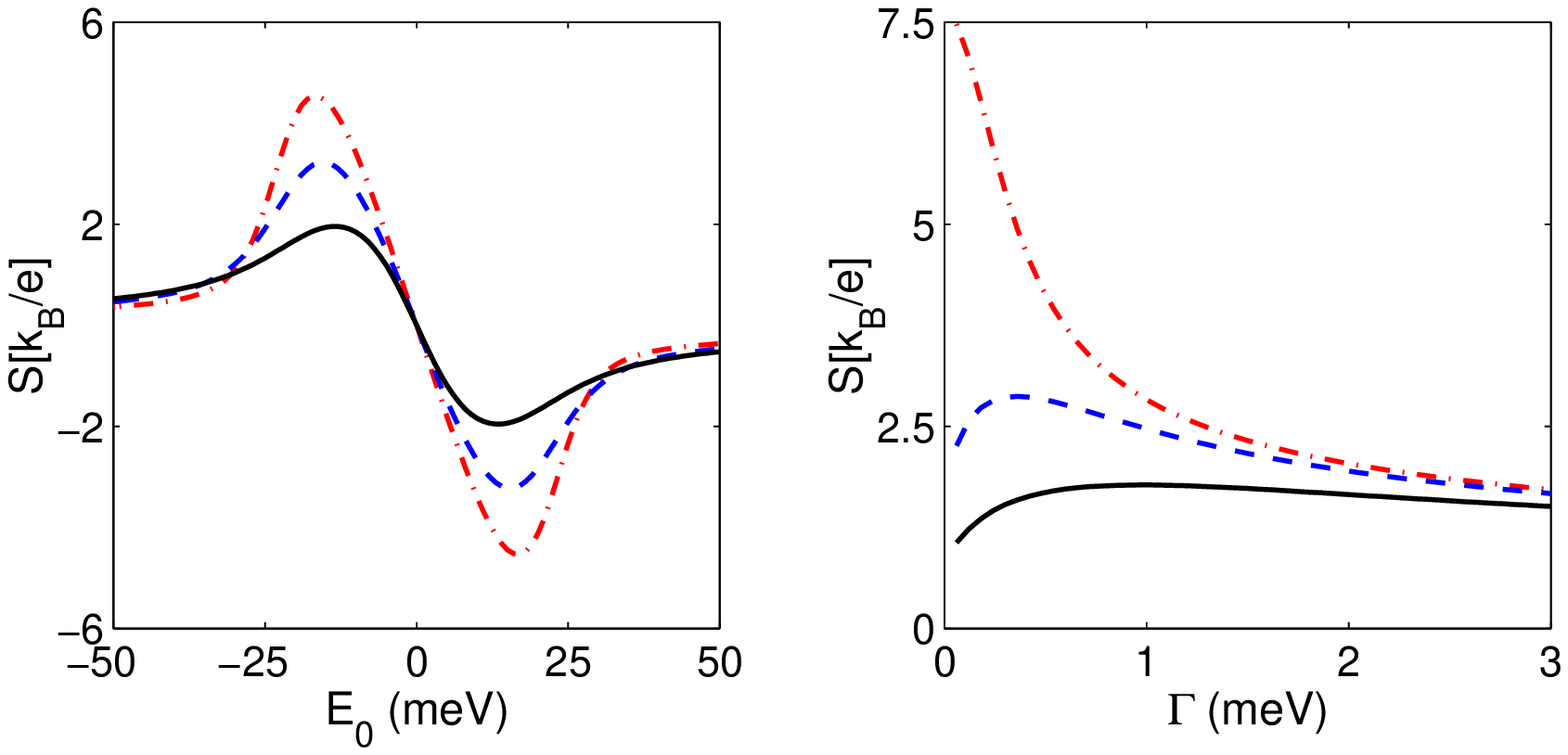}
\includegraphics[width=8.8cm,height=4.5cm]{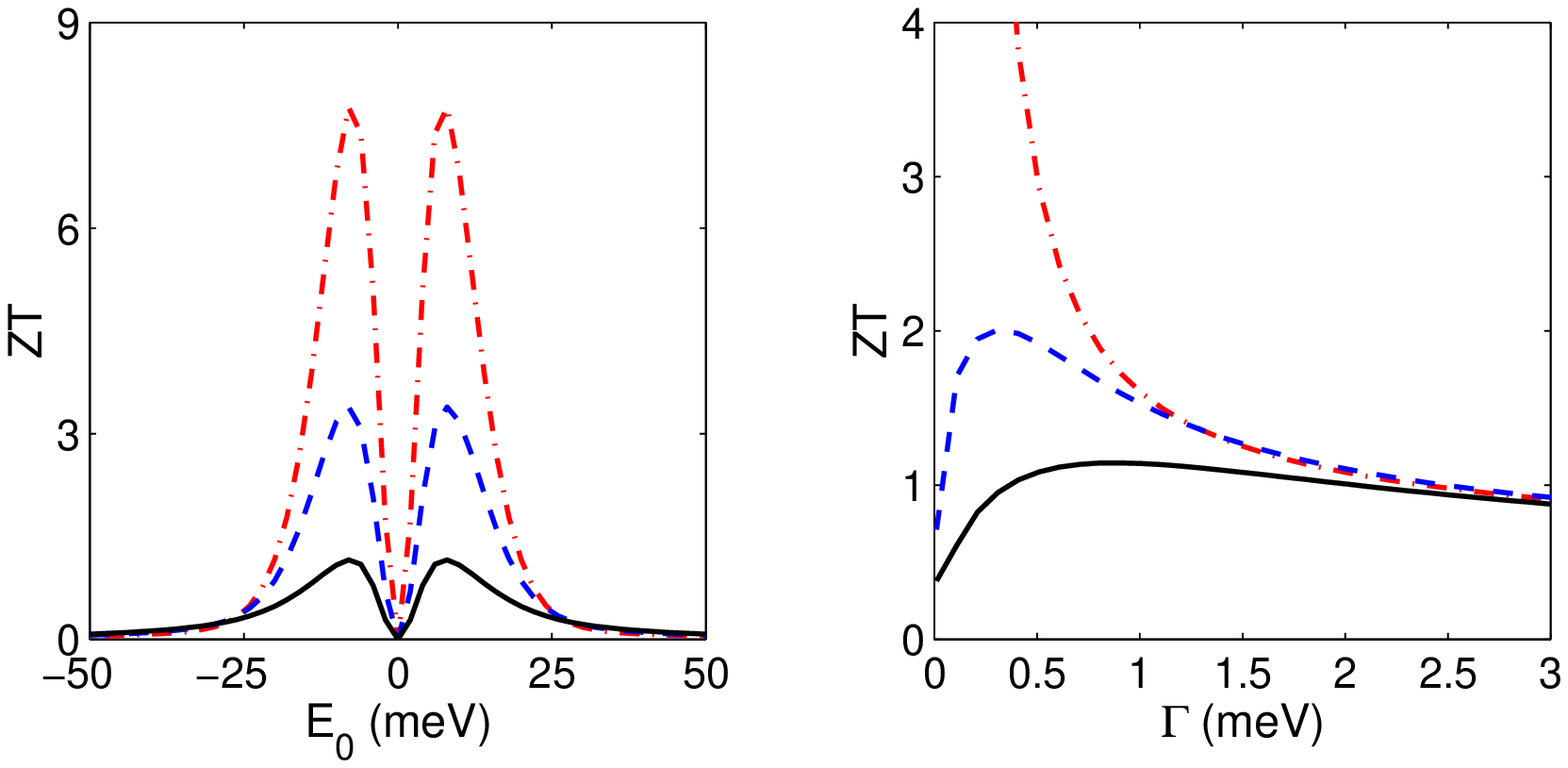}
\caption{(Color online) Thermopower and figure of merit as functions of the bridge  level position  (left panels) and of the parameter $ \Gamma $ characterizing the coupling between the bridge and the electrodes (right panels). The curves are plotted at $ k_BT = 2.6 meV, $   $ \lambda = 0 $ (dash-dotted lines), $ \lambda = 0.25 meV $ (dashed lines), and $ \lambda = 1meV $ (solid lines) assuming $ \Gamma = 1.25 meV $ (left panels) and $ E_0 = 10 meV $ (right panels).
}
 \label{rateI}
\end{center}\end{figure}

It was first shown by Sofo and Mahan \cite{51} and then confirmed in several later works (see e.g. Ref. \cite{21}) 
  that the figure of merit diverges when $ \Gamma $ approaches zero provided that the system is characterized by zero phonon contribution to the thermal conductance, and the effects of Coulomb interactions between electrons on the bridge are disregarded. The results obtained in the present work agree with this conclusion. In the right bottom panel of the Fig. 3, the divergence of $ ZT $ within the limit $ \Gamma \to 0 $ in the absence of the electron-phonon interactions is clearly illustrated. The junction figure of merit is limited due to the effect of thermal phonons  associated with the electrodes. The stronger these interactions are, the lower in magnitude maximum values of both $ ZT $ and thermopower become. Comparing the present results with those reported in Ref. \cite{21} one may presume that the thermal phonons take the part of phonon thermal conductance (which equals zero for the considered system) in limiting the maximum value of $ ZT $ and removing the divergence. Unlike $ ZT ,$ the thermopower remains finite at small values of $ \Gamma $ even when the electron-phonon interactions are disregarded, as illustrated in the Fig. 3 (see right top panel). This means that the divergence of $ ZT $ originates from extremely strong violation of the Wiedemann-Franz law resulting in the divergence of the Lorentz ratio.

\begin{figure}[t] %%% fig. 4
\begin{center}
\includegraphics[width=8.8cm,height=4.5cm]{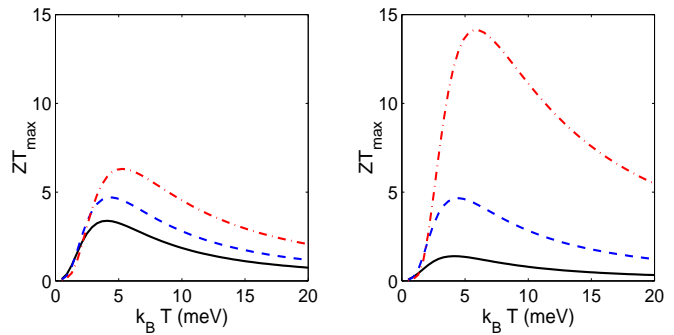}
\caption{(Color online) Maximum value of $ ZT $ as a function of temperature. The curves shown in the left panel are plotted assuming $ \lambda = 0.25meV,\ \Gamma = 0.3 meV $ (solid line), $ \Gamma = 0.7meV $ ( dashed line), $ \Gamma =1.25 meV $ (dash-dotted line). In the right panel, the curves are plotted at $ \Gamma = 0.7 meV, \ \lambda = 0 $ (dash-dotted line), $ \lambda = 0.25meV $ (dashed line), $ \lambda = 1meV $ (solid line). 
}
 \label{rateI}
\end{center}\end{figure}

In analyzing temperature dependencies of  thermoelectric characteristics  of systems consisting of a  molecule/quantum dot linking two electrodes, it was established that usually the figure of merit $ ZT $ is a nonmonotonous function of temperature (see e.g. Refs. \cite{9,19,20,21,25,46}. Also, the present results show that at low temperatures $ ZT $ increases as the temperature enhances and it reaches a maximum value at certain temperature $ T_0. $ As $ T $ further  rises, the figure of merit decreases approaching zero when the temperature significantly exceeds $ T_0. $ This is illustrated in the Fig. 4. The value of the optimal temperature $ T_0 $ as well as the corresponding value of $ ZT_{max} $ is determined by the relation between the coupling energies $ \lambda $ and $ \Gamma. $ Assuming that $ \lambda $ is fixed, one observes that $ ZT_{max} $ takes on greater values and the optimal temperature $ T_0 $ becomes higher as $ \Gamma $ increases. On the contrary, enhancement of $ \lambda $ at fixed $ \Gamma $ leads to a significant decrease of $ ZT_{max}, $ and shifts $ T_0 $ to a lower value.  Thus the electron interactions with the thermal baths  suppress $ ZT $ values. Molecular vibrations may affect thermoelectric efficiency of the considered nanoscale systems in a similar way, as discussed in several works (see e.g. Refs. \cite{13,21,30}).

The character  of temperature dependence of $ ZT $ displayed in the Fig. 4 indicates that while the transition from coherent and elastic tunneling to the dissipative transport significantly reduces $ZT $ values, the general character of temperature dependence of the figure of merit remains unchanged.
 At low temperatures, erosion of the sharp step in the Fermi distribution functions for the electrodes occurring at $ E  = \mu $ creates better opportunities for the electron tunneling through the system. However, when the temperature exceeds a certain value, the same process starts to hinder electron transport. Also, at sufficiently strong electron-phonon interactions, the peak value of the electron transmission decreases   
  bringing further reduction of the thermoelectric efficiency.

Although considerable efforts are applied to reach understanding of combined effects of electron-electron and electron-phonon interactions on the thermoelectric transport, this subject is not fully investigated so far. Now, we reconsider the above results taking into account previously disregarded Coulomb interactions between electrons on the bridge of a thermoelectric junction.  
 Then the expression (\ref{15}) for the electron Green's function may be employed to compute the scattering probabilities and, ultimately, the electron transmission $ T(E) $ and measurable characteristics of thermoelectric transport. In further analysis we assume that the linker (molecule/quantum dot) is weakly coupled to the electrodes so that the charging  energy  $ U $  significantly exceeds the coupling parameter $ \Gamma. $   The results of these  computations are displayed in the Figs. 5,6.

\begin{figure}[t] %%% fig. 5
\begin{center}
\includegraphics[width=8.8cm,height=4.5cm]{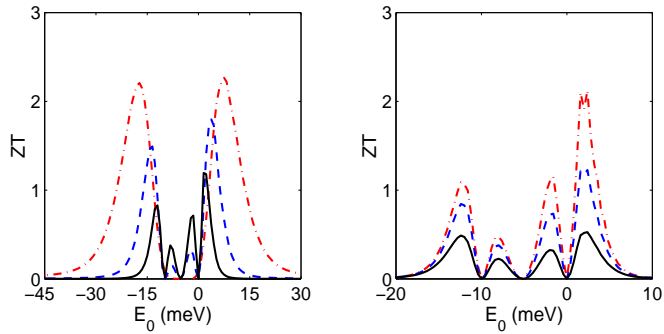}
\caption{(Color online) Combined effect of electron-electron and electron-phonon interactions  on the dependence of $ ZT $  of the bridge level position. The curves are plotted at $ U = 10meV.$ Left panel: $ \Gamma = 1,25 meV,\ \lambda = 0.25 meV ,$  $ k_BT = 2.6 meV $ (dash-dotted line), $ k_BT = 1.3 meV $ (dashed line), $ k_BT = 0.7 meV $ (solid line). Right panel: $ \Gamma = 1.25meV,\ k_BT = 0.7meV,$  $ \lambda = 0  $ (dash-dotted line), $ \lambda = 0.25 meV $ (dashed line), $ \lambda = 1meV $ (solid line).
}
 \label{rateI}
\end{center}\end{figure}

As shown in the Fig. 5, the dependence of $ ZT $ of  $ E_0 $ undergoes significant changes as the temperature increases. At low temperatures, $ ZT $ exhibits two pairs of peaks of unequal height situated near $ E_0 = \mu $ and $ E_0 = \mu - U,$ respectively. At higher temperatures two peaks making a pair cling together so that each pair is transformed to a single peak. At sufficiently high temperatures, these peaks become nearly equal in height, and their tops are shifted farther away from each other. The  curves displayed in the left panel of the Fig. 5  are plotted assuming that temperature is noticeably lower than the temperature $ T_0 $ providing the maximum value of $ ZT ,$ as shown in the Fig. 4. We cannot explicitly compare the results represented in these figures  because the curves shown in the Fig. 4 are plotted disregarding electron-electron interactions. However, we may conjecture that further increase of temperature accompanied by intensification of scattering processes will bring furthermost rise of $ ZT $ peaks as well as it happens in the case when one neglects electron-electron interactions. Also, we may expect that the increase in the peaks heights would be replaced by their reduction as the temperature would exceed a certain value. An explicit effect of electron-phonon interactions on the figure of merit is shown in the right panel of the Fig. 5. Again, one may observe the suppression of $ ZT $ originating from these interactions.

\begin{figure}[t] %%% fig. 6
\begin{center}
\includegraphics[width=8.8cm,height=4.5cm]{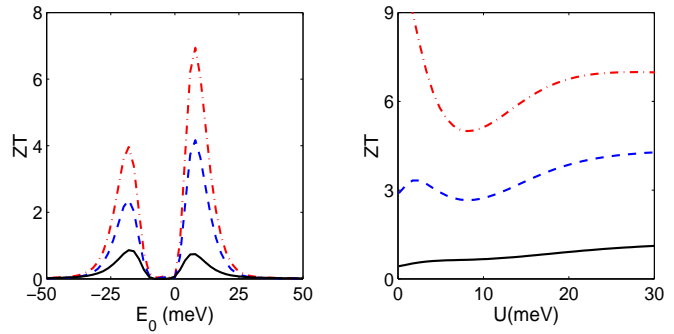}
\caption{(Color online) Left panel: Combined effect of electron-electron and electron-phonon interactions on the dependence of ZT of the level position at higher temperatures. Right panel: maximum value of $ ZT $ as function of charging energy. Curves are plotted assuming $ k_BT = 2.6meV,\ \Gamma = 1.25meV,\ U = 10 meV $ (left panel) $ \lambda = 0 $ (dash-dotted lines), $ \lambda = 0.25 meV $ (dashed lines), $ \lambda = 1meV $ (solid lines). 
}
 \label{rateI}
\end{center}\end{figure}

Further  illustration of the influence of thermal phonons on the figure of merit is presented in the Fig. 6. The curves shown in the left panel of this figure are plotted at a moderately high temperature $(k_B T = 2.6 meV) $ when the adjacent peaks are already merged, so that  $ ZT $ exhibit only two maxima. Omitting electron-phonon interactions, one observes a significant difference in the peaks heights. This difference originates from the characteristic features of electron density of states on the bridge level which are manifested in the characteristics of coherent electron transport. As the  electrons interaction with thermal phonons strengthens, the peaks heights become leveled. At fixed temperature, maximum value of $ ZT $ is determined by the relation between the charging energy $ U, $ and coupling strengths $ \lambda $ and $ \Gamma. $ We remark that the suppression of $ ZT $ due to electron-phonon interactions may be replaced by its promotion which occurs due to a combined effect of electron-electron and electron-phonon interactions \cite{33}. However, this increase of thermoelectric efficiency is expected to appear when electron-phonon interactions and Coulomb repulsion between electrons are comparable in strength. These conditions are different from those considered in the present work.
 Disregarding  for a while the effect of phonons, one observes that $ ZT $ takes on greater values within the limits of low $(U \ll k_B T)$ and high $ (U \gg k_BT)$ values of charging energy, and it drops at intermediate values of $ U. $ This behavior was previously described and explained within the sequential hopping approximation for the electron transmission through a thermoelectric junction \cite{21}. Electron-phonon interactions promote washing out of these features.

\section{iv. Conclusion} 

In conclusion, we remark that thorough studies of thermoelectric properties of nanoscale systems  taking into account both electron and phonon transport as well as diverse effects arising due to electron-electron and electron-phonon interactions are not completed so far. In several earlier works this theoretical research was carried out employing single-particle scattering approach pioneered by Landauer in the context of charge transport in mesoscopic and nanoscale systems. These ideas were generalized to phonon transport through nanoscale junctions 
\cite{30,32,42,52,53}. Within this approach, transport characteristics of a considered nanoscale system are expressed in terms of electron and phonon transmission functions. The latter were computed employing several methods, including some based on scattering matrices formalism \cite{42,54}. Later, these methods were mostly abandoned in favor of more advanced formalisms such as NEGF and/or various modifications of quantum rate equations. However, potential usefulness of the approaches based on scattering theory is not exhausted so far. 

These approaches have an advantage of being computationally simple and less time and effort consuming than advanced formalisms. At the same time, their shortcomings  could be largely removed by incorporating some NEGF based results into a computational scheme. In the present work we suggest such approach, and we employ it to theoretically analyze some effects of electron-phonon interactions on the efficiency of nanoscale thermoelectric junctions. Presently, various manifestations of electron-phonon interactions in thermoelectric transport characteristics of nanoscale molecular junctions are already explored, and the research is still going on. However, the research efforts were and still are mostly concentrated on the effects arising from vibrational modes on the molecules linking the electrodes. Less attention was paid to the influence of thermal phonons associated with random nuclear motions in the ambience. Here, we focus on the analysis of thermal phonons on the electron transport. We show that direct interaction of electrons with thermal phonons assuming that  these phonons are assembled in two baths associated with the electrodes.
may significantly affect thermoelectric efficiency of molecular junctions and similar nanoscale systems. 

Specifically, we show that electron-phonon interactions assist the increase of the scattering probabilities thus destroying the coherence of electron transport and promoting energy dissipation. When the electron-phonon coupling  becomes sufficiently strong, this brings a significant suppression of both thermopower and thermoelectric figure of merit thus worsening thermoelectric efficiency of a considered system. This effect is illustrated in the Figs. 3,4. We remark that $ \lambda $ and thermal energy $ k_BT $ appear as cofactors in the expression for the scattering probability $ \epsilon $ (see Eq. (\ref{29})), so they affect it in a similar way. When either $ \lambda $ or $ k_BT $ increases, this results in strengthening of dephasing in the electron transport. However, entire effects of these two parameters on the thermoelectric properties of considered systems are unidentical.
While the strengthening of electron-phonon interactions always leads to reduction of $ ZT, $ the rise of temperature can promote the figure of merit increase provided that temperature does not exceed a certain value. This may be explained by the fact that besides affecting the intensity of scattering, the temperature influences distributions of electrons in the electrodes, and it may either assist or hinder their transport through the system. Also, we analyzed the combined effect of electron-electron and electron-phonon interactions on thermoelectric properties. Obtained results agree with those reported in the earlier works \cite{21,33}. In particular, it was confirmed that $ ZT $ exhibits a minimum at a certain value of the charging energy $ U $ which becomes less distinct at stronger values of $ \lambda. $

The suggested computational scheme may be generalized to include vibrational modes.
For this purpose, one may add to the adopted model simulating a thermoelectric junction an extra reservoir representing vibrons. Also, one may mimic the molecular bridge by a set of energy levels thus opening the way to studies of interference effects.  For a realistic molecular junction, relevant energies may be computed  using either density functional theory or other method of electronic structure calculations. Finally, the proposed scheme may appear helpful in studies of thermoelectric transport beyond linear regime in temperature. On these grounds, we believe that presented method and results could help to reach better understanding of some important aspects of thermal transport in molecular junctions and similar nanoscale systems.
\vspace{2mm}

 {\bf Acknowledgments:}
 This work was  supported  by  NSF-DMR-PREM 0934195 and NSF-EPS-1010094. The author  thank  G. M. Zimbovsky for help with the manuscript.

\end{document}